\documentclass[jkps,twocolumn,fleqn,showkeys]{revtex4-2}
\usepackage{amssymb}
\usepackage{amsmath}
\usepackage{graphicx}
\usepackage{bm,multirow}
\usepackage[]{color}

\begin{document}
\title{Unconventional Thermal Expansion in quasi-one-dimensional monoclinic BaIrO$_3$}


\author{Jinwon Jeong, Bin Chang, Han-Jin Noh}
\email{ffnhj@jnu.ac.kr}
\affiliation{Department of Physics \& Center for Quantum Technologies, Chonnam National University, Gwangju 61186, Korea}
\author{Seongsu Lee}
\affiliation{Neutron Science Division, Korea Atomic Energy Research Institute, Daejeon 34057, Korea}


\date{\today}

\begin{abstract}
We have investigated the temperature dependence of the crystal structure of quasi-one-dimensional monoclinic BaIrO$_3$ using X-ray diffraction.
Diffraction patterns were measured across a temperature range from 13 K to 300 K, with 5-degree steps, and Rietveld refinements were performed to extract the relevant lattice parameters. 
The resulting cell volumes exhibit a significant deviation from the Debye model predictions for lattice-specific heat within a reasonable range of the Debye temperature, Gr{\"u}neisen parameter, and bulk modulus. 
This suggests an invar-like, unconventional thermal expansion behavior. 
The deviation begins near the weak ferromagnetic transition temperature, indicating a strong correlation with changes in the electronic and magnetic structure of monoclinic BaIrO$_3$.
\end{abstract}

\keywords{X-ray diffraction, BaIrO$_3$, invar effect, weak ferromagnetism}

\maketitle

\begin{figure}
\includegraphics[width=7.0 cm]{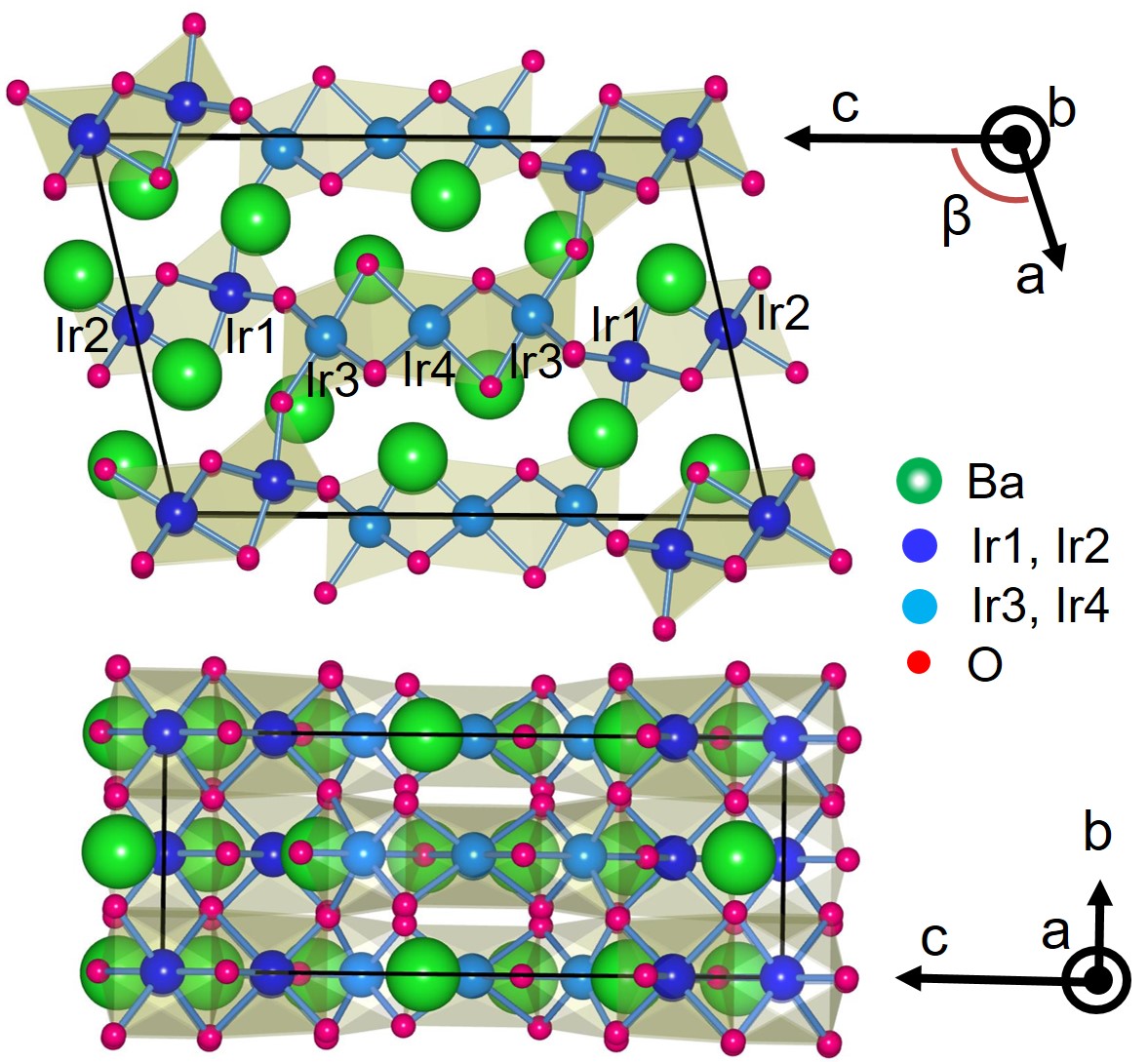}
\caption{\label{fig1}
Crystal structure of 9R-type monoclinic (C2/m) BaIrO$_3$. Upper panel is \textbf{b}-axis view and lower \textbf{a}-axis view. Ir1 - Ir4 denote crystallographically distinct Ir sites in a unit cell. Solid parallelogram denotes the unit cell.
}
\end{figure}

Quasi-one-dimensional monoclinic BaIrO$_3$ has long been studied due to its exotic physical properties, though a comprehensive understanding remains unclear. 
With respect to the crystal structure phases of BaIrO$_3$, five polytypes - 9R, 4H, 5H, 6H, and 3C - have been identified through pressure-dependent diffraction studies, following earlier trial-and-error efforts. \cite{Siegrist, Cheng1, Cheng2}
Among these, the most stable phase at ambient pressure is so-called 9R-BaIrO$_3$.
This polytype consists of quasi-one-dimensional Ir$_3$O$_{12}$ trimers linked by corner-sharing oxygen atoms as depicted in Fig.~\ref{fig1}.

The electronic and magnetic ground states of this iridate are both complex and exotic.
Resistivity vs. temperature curves on both polycrystalline and single-crystal samples reveal three anomaly points T$_{c1}$=175 K, T$_{c2}$$\sim$80 K, and T$_{c3}$=26 K with significant anisotropy between c-axis and ab-plane directions.\cite{Cao1, Cao2, Cheng1, Terasaki}
Special attention has been focused on the highest temperature point (T$_{c1}$) among them as it is associated with both an electronic and a magnetic phase transition.
Regarding the electronic phase, the charge density wave (CDW) scenario was proposed based on the resistivity measurements and optical spectra.\cite{Cao1}
However, the absence of superstructure peaks in diffraction patterns and the negative slope above T$_{c1}$ in the resistivity curves challenge this model.

The magnetic phase remains unclear as well.
While canted ferromagnetism, based on a local spin 1/2 model in an Ir$_3$O$_{12}$ trimer, has been suggested to explain the weak magnetic moment, both band structure calculations and X-ray Magnetic Circular Dichroim (XMCD) measurements indicate that the magnetism is more closely related to the spins and orbitals of the conducting Ir 5$d$ electrons.\cite{Maiti, Ju, Laguna-Marco}
In this context, an effective total angular momentum state formed by the strong spin-orbit interaction (SOI) in 5$d$ transition metal elements plays a key role in explaining the high temperature insulating phase.\cite{Kim1, Kim2}
Recognizing the importance of the SOI in this system, Terasaki {\it et al.}  proposed a new model that it is a Mott insulator to charge order insulator transition, based on their transport, magnetization and X-ray diffraction measurements on Ru-doped BaIrO$_3$ samples.\cite{Terasaki}
More recently, Chang {\it et al.} also proposed a similar model, derived from their temperature-dependent neutron diffraction analysis, in which an effective $J_{\rm{eff},1/2}$ Mott insulator competes with a charge gap phase.\cite{Chang}

In this study, we examine the temperature dependence of the crystal structure of quasi-one-dimensional monoclinic 9R-BaIrO$_3$ using X-ray diffraction.
While a crystal structure analysis of each phase using neutron diffraction was previously reported in our earlier work, the data were insufficient to reveal detailed dependencies due to the coarse temperature steps employed.\cite{Chang}
To address this, we conducted X-ray diffraction (XRD) measurements across a temperature range from 13 K to 300 K with 5-degree increments, and performed Rietveld refinements to extract the relevant lattice parameters.
This approach allowed us to identify an invar-like, unconventional thermal expansion behavior in this iridate.
The resulting cell volumes exhibit a significant deviation from the Debye model predictions for lattice-specific heat within a reasonable range of the Debye temperature, Gr{\"u}neisen parameter, and bulk modulus. 
The deviation begins near the weak ferromagnetic transition temperature, indicating a strong correlation with changes in the electronic and magnetic structure of monoclinic BaIrO$_3$.

The polycrystalline powder samples of BaIrO$_3$, sourced from the same sample batch as in our previous work, were used for X-ray diffraction and transport measurements.\cite{Chang}
Low temperature XRD measurements were conducted using a commercial diffractometer (Empyrean) equipped with a closed-cycle helium cryostat (Oxford PheniX).
Resistivity vs. temperature measurements were carried out using a standard four-probe method with a custom-made transport measurement system.

\begin{figure}
\includegraphics[width=8.5 cm]{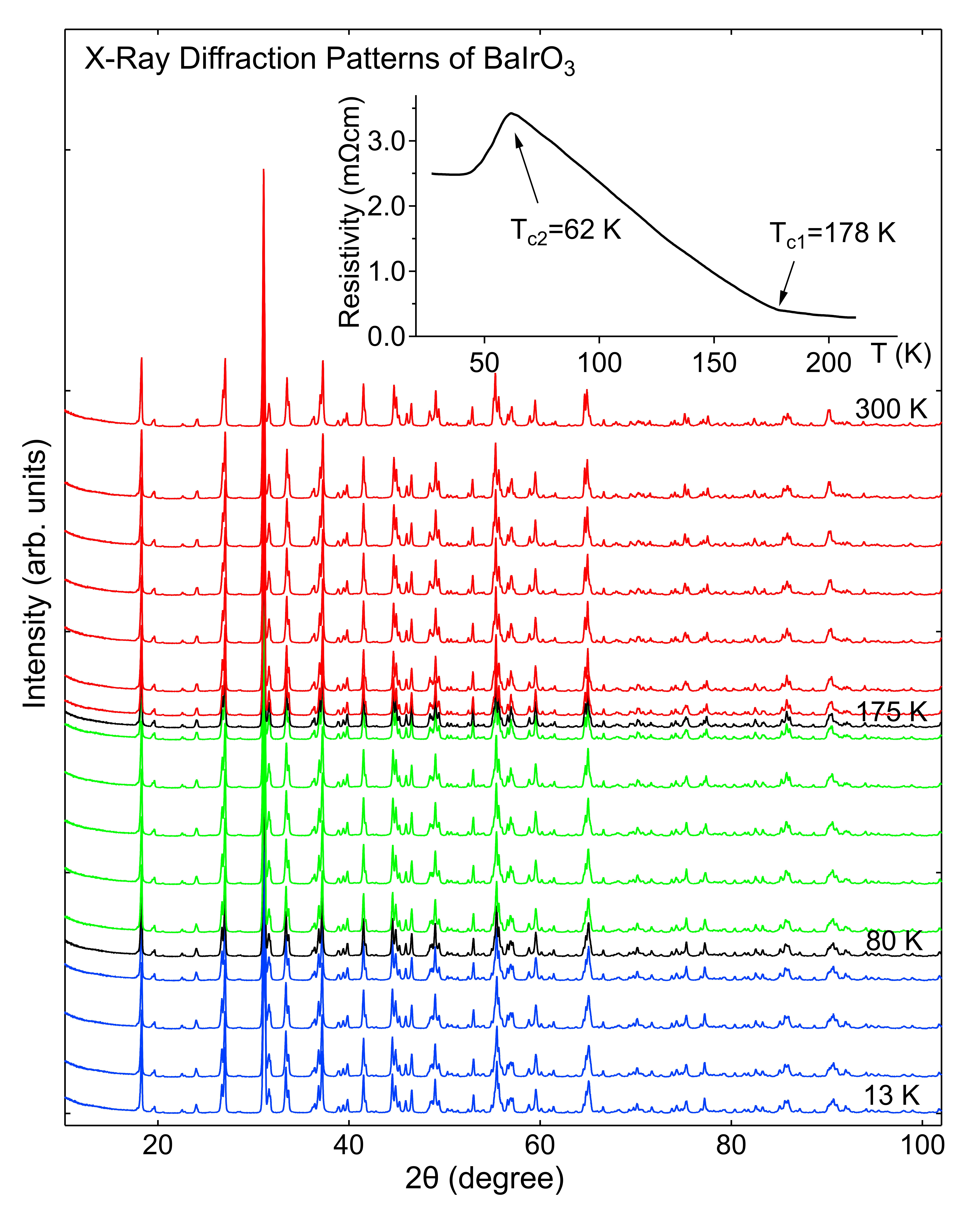}
\caption{\label{fig2}
X-ray diffraction patterns of monoclinic BaIrO$_3$ at the temperature range from 13 to 300 K. 
The data were obtained at every 5 K step, but displayed at $\sim$20 K step for clarity. 
The patterns for phase boundary are displayed in black color.  
(Inset) Resistivity vs. temperature curve.
}
\end{figure}

Figure~\ref{fig1} illustrates the unit cell of the monoclinic BaIrO$_3$ crystal structure, which belongs to the space group $C2/m$.
This polytype is conventionally referred to as ''9R-type'' BaIrO$_3$, named after 9R-type BaRuO$_3$ without monoclinic distortion.\cite{Cheng2}
However, strictly speaking, it does not adhere to the standard polytype nomenclature.\cite{Guinier}
Crystallographically, the unit cell contains four independent Ir sites (Ir1 - Ir4), six oxygen sites, and two types of Ir trimers (Ir1-Ir2-Ir1 and Ir3-Ir4-Ir3).
The XRD patterns were measured over a temperature range from 13 to 300 K with 5 K increments, but for clarity, only a reduced set of data is presented in Fig.~\ref{fig2}.
As in our earlier neutron diffraction study, no super-structure peaks are detected across the phase boundaries.\cite{Chang}
To check the transport properties of our samples, we also measured resistance vs. temperature curve as shown in the inset of Fig.~\ref{fig2}, which exhibits behavior consistent with previously reported data.\cite{Cheng1,Cao1,Cao2,Terasaki}
It is noteworthy that 9R-BaRuO$_3$ exhibits a similar partial-gap openning in the optical spectra below the metal-insulator transition temperature T$^{*} \sim$140 K without undergoing a structural transition.\cite{Nam}
The common CDW-like behaviors without super-structure formation in these 9R polytype transition metal oxides suggest a strong connection to an instability in their quasi-one-dimensional structure.

\begin{figure}
\includegraphics[width=8.5 cm]{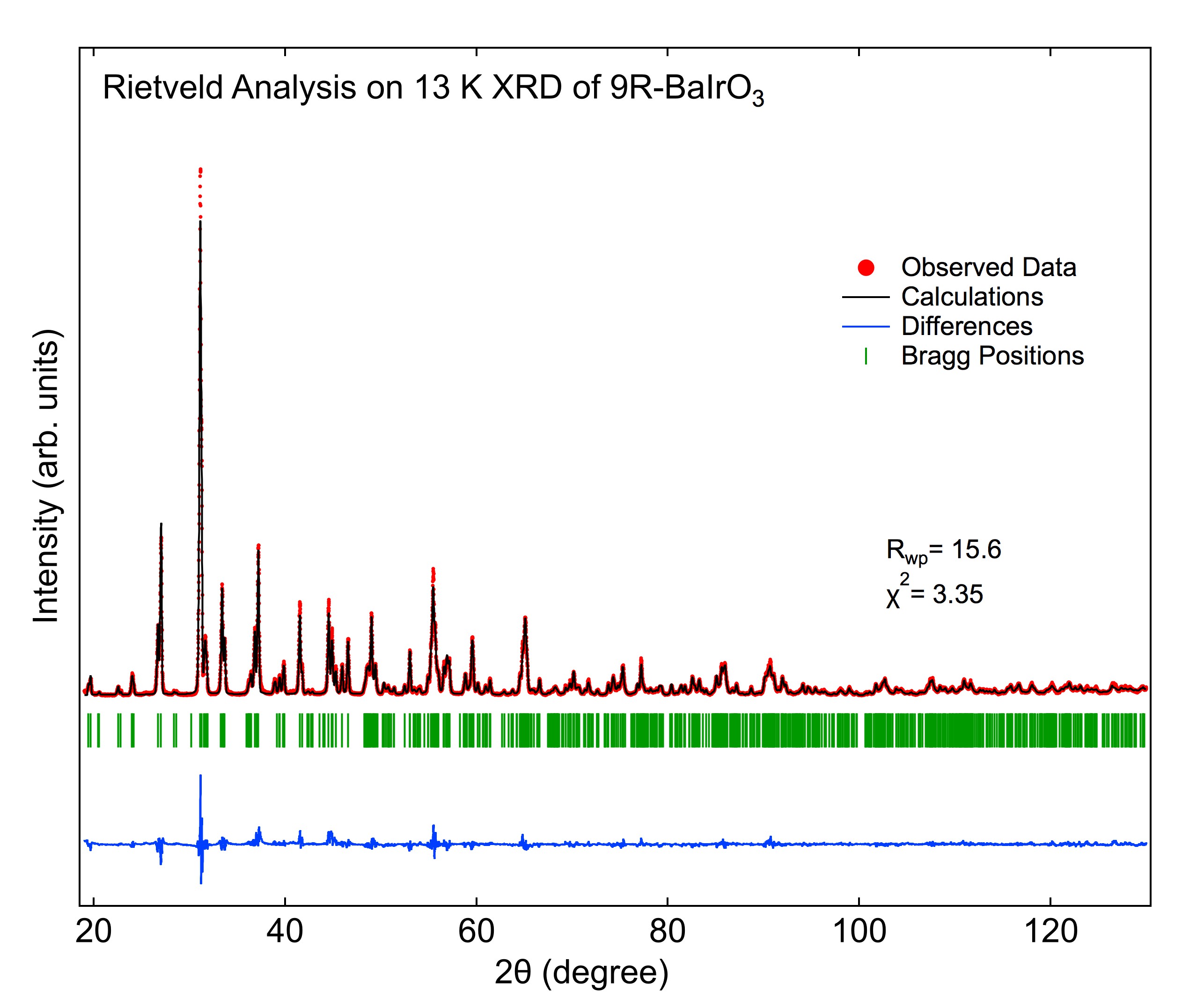}
\caption{\label{fig3}
Rietveld refinement result for the 13 K XRD pattern of BaIrO$_3$.
}
\end{figure}

Using these XRD data, we conducted Rietveld refinements with Fullprof 2000 to scrutinize the temperature dependence of the structure.\cite{Roisnel}
For simplicity, we present the refinement results only for the 13 K data in Fig.~\ref{fig3}.
The fitting yielded goodness-of-fit parameters, including a weighted profile R-factor R$_{wp}$=15.6 and a chi-square value $\chi^{2}$=3.35.
The parameters for the other temperature data are of similar magnitude.
All our XRD patterns are successfully reproduced by adjusting only the cell parameters and atom positions under the monoclinic C${2/m}$ structure and are consistent with the result of the neutron diffraction study reported in our earlier work.\cite{Chang}
This can help to rule out a possible misleading that may be induced from the low cross-sections of X-ray to light elements such as oxygen.

\begin{figure}
\includegraphics[width=8.5 cm]{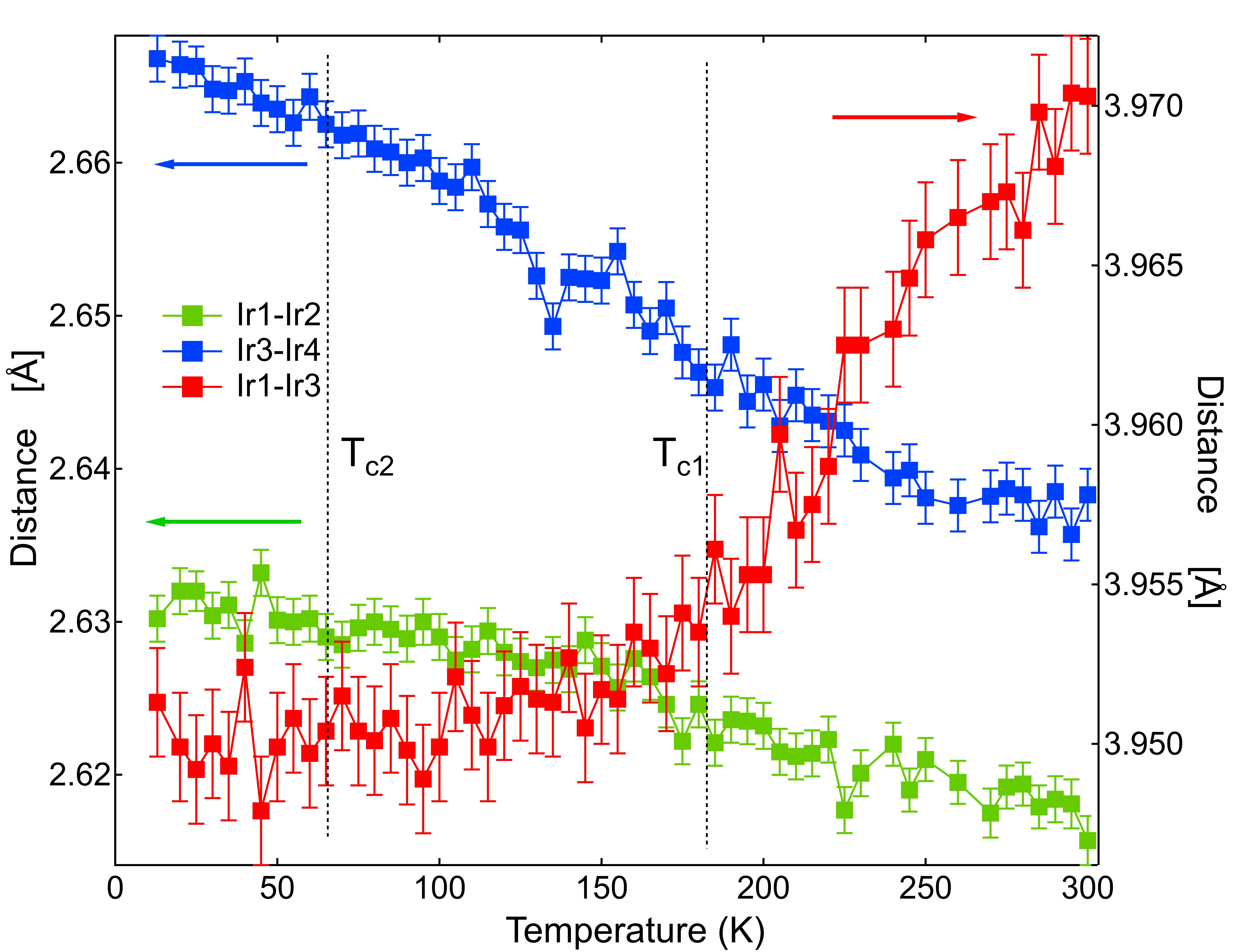}
\caption{\label{fig4}
Temperature dependence of the Ir-Ir cation distances for each bonding type, based on the structure refinements.
}
\end{figure}

To figure out the structural evolution of the iridate with temperature, we calculated the distances between all types of Ir-Ir cations from the refinement results and plotted them as a function of temperature, as shown in Fig.~\ref{fig4}. 
Considering the linkage of the two types of Ir$_3$O$_{12}$ trimers (Ir1-Ir2-Ir1 and Ir3-Ir4-Ir3) through corner-sharing of oxygen atoms, it means that the length of each type of trimer decreases with increasing temperature, while the distance between trimers increases.
Notably, the inter-trimer distance begins to increase abruptly around T$_{c1}$ with rising temperature, which is consistent with the competition model between the high temperature effective $J_{\rm{eff},1/2}$ Mott insulating phase and the low temperature charge gap phase based on the trimer atom picture.\cite{Terasaki, Chang}

\begin{figure}
\includegraphics[width=8.5 cm]{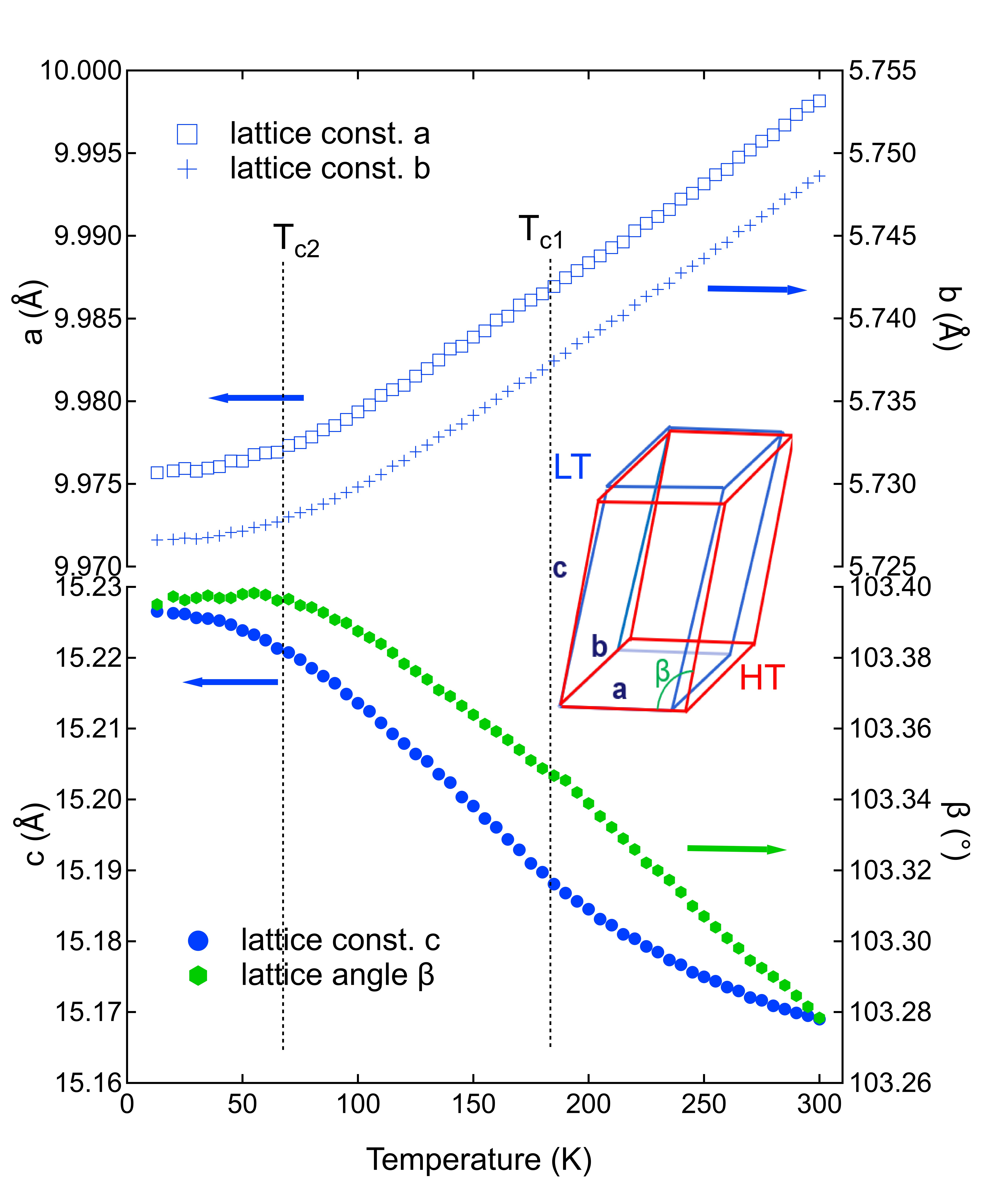}
\caption{\label{fig5}
Temperature dependence of the lattice constants and the lattice parameter angle of monoclinic BaIrO$_3$.
(Inset) Schematic monoclinic unit cell for high/low temperature range.
}
\end{figure}

Figure~\ref{fig5} displays the thermal expansion behavior of the cell parameters for 9R-BaIrO$_3$.
Interestingly, the lattice constant \textbf{c} decreases with increasing temperature, exhibiting a concave feature around T$_{c1}$.
The lattice parameter angle $\beta$ also decreases with rising temperature, displaying a cusp-like feature around T$_{c1}$.
Meanwhile, the lattice constants \textbf{a} and \textbf{b} show typical thermal expansion behaviors.
A schematic representation of the monoclinic unit cell at a high (low) temperature is shown in the inset of Fig.~\ref{fig5}, with a red (blue) parallelogram, highlighting the temperature dependence.

\begin{figure}
\includegraphics[width=8.5 cm]{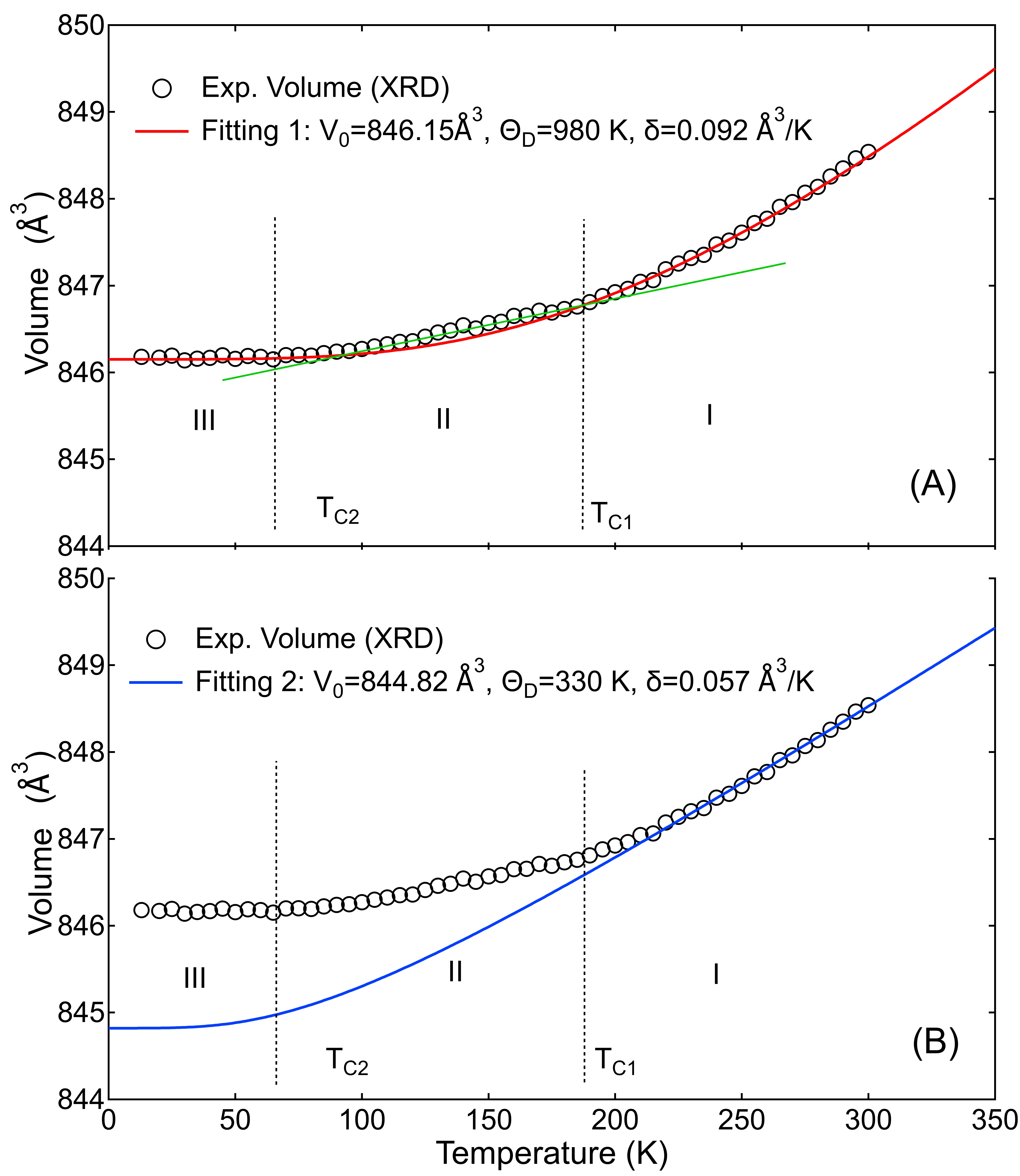}
\caption{\label{fig6}
Temperature dependence of the unit cell volume of monoclinic BaIrO$_3$ calculated from the XRD refinement results.
(A) The fitting curve (red line) was obtained using the Debye temperature as a free fitting parameter.
(B) The fitting curve (blue line) was obtained using the experimental Debye temperature in Ref\cite{Cheng1}.
}
\end{figure}

The unit cell volumes were calculated using these lattice parameters, as shown in Fig.~\ref{fig6}.
The curve is roughly divided into three regions.
From room temperature to approximately T$_{c1}$ (region I), the curve exhibits a slight concavity.
Between T$_{c1}$ and T$_{c2}$ (region II), the curve appears to form a straight line as indicated by the green line in Fig.~\ref{fig6}(A).
Below T$_{c2}$ (region III), the slope approaches zero.
In order to analyze the thermal volume expansion behavior more systematically, we employed the Debye model, which accounts for the anharmonic effects of lattice vibrations contributing to thermal expansion in solids.\cite{Ashcroft}
Based on this model, the thermal volume expansion behavior is given by the following formula\cite{Kiyama}:
$$ \begin{array}{ll} V(T) & \cong V_0 + \int_{0} ^{T} \frac{\gamma C_v}{B} dT \\
  & \cong V_0 + \frac{9\gamma Nk_B T}{B}(\frac{T}{\Theta_D})^3 \int_{0} ^{\Theta_D/T} \frac{x^3}{e^x -1} dx. \end{array}$$
Here, $V_0$, $\gamma$, $\Theta_D$, and $B$ represent the volume at 0 K, the Gr\"{u}neisen parameter, the Debye temperature, and the bulk modulus, respectively.

First, we attempted to fit the volume curve using the Debye model, with $V_0$, $\Theta_D$, and $\delta\equiv 9\gamma{N}k_BT/B$ as free fitting parameters.
The resulting red curve in Fig.~\ref{fig6}(A), obtained with $V_0$=846.15 \AA$^3$, $\delta$=0.057, and $\Theta_D$=980 K, provides a reasonable fit, except for the deviations in region II.
However, this fitting presents a significant issue: the Debye temperature is too high.
The literature value, reported as 268.2 K, was derived from specific heat measurements.\cite{Cheng1} 
When the Debye temperature is fixed to this reported value, we can only partially fit the model curve (blue line) to the experimental volume curve above T$_{c1}$, as shown in Fig.~\ref{fig6}(B).
Within a physically reasonable parameter space, a single Debye model function fails to fully reproduce the experimental volume expansion curve. 

The results of this analysis strongly suggest that the thermal volume expansion behavior of 9R-BaIrO$_3$ deviates from conventional types.
Although the thermal volume expansion coefficient is not exactly zero, it closely resembles an invar-type behavior.
The invar effect in oxides has been rarely reported, with the only notable example being SrRuO$_3$, an itinerant electron ferromagnetic oxide with a T$_c$$\sim$160 K.\cite{Kiyama}
A key common feature between 9R-BaIrO$_3$ and SrRuO$_3$ is that both exhibit invar behavior below their respective magnetic phase transition temperatures. 
While a concensus on the microscopic origin of the invar effect has yet to be reached, it is generally agreed that magnetic orderings compensate for the thermal expansion caused by lattice vibrations.\cite{Matsui, Ehn}
A notable difference between the two oxides also lies in the lattice thermal expansion.
9R-BaIrO$_3$ shows negative thermal expansion only in \textbf{c} direction, as seen in Fig.~\ref{fig5}, whereas SrRuO$_3$ exhibits negative thermal expansion in all directions.\cite{Kiyama}
This can be reasonably atrributed to the pseudo-one-dimensional structure of 9R-BaIrO$_3$.
The directional anisotropy in thermal expansion makes this iridate truly unique among invar materials.
To the best of our knowledge, this is the first invar system with a low dimensional structure, and it is expected to offer valuable insights for the microscopic understanding of the invar effect.
 
To summarize, we have investigated the temperature dependence of the crystal structure of quasi-one-dimensional monoclinic BaIrO$_3$ using XRD.
Diffraction patterns were measured across a temperature range from 13 K to 300 K, with 5-degree steps, and Rietveld refinements were performed to extract the relevant lattice parameters. 
The resulting cell volumes exhibit a significant deviation from the Debye model predictions for lattice-specific heat within a reasonable range of the Debye temperature, Gr{\"u}neisen parameter, and bulk modulus. 
This suggests an invar-like, unconventional thermal expansion behavior. 
The deviation begins near the weak ferromagnetic transition temperature, indicating a strong correlation with changes in the electronic and magnetic structure of monoclinic BaIrO$_3$.

\acknowledgments
This work was supported by the National Research Foundation (NRF) of Korea Grant funded by the Korean Government (MEST) (No. 2019R1I1A3A0105818813).

\end{document}